\documentclass[10pt, conference, letterpaper]{IEEEtran}
\IEEEoverridecommandlockouts
\usepackage{cite}
\usepackage{amsmath,amssymb,amsfonts}
\usepackage{algorithmic}
\usepackage{graphicx}
\usepackage{textcomp}
\usepackage{graphicx}   
\usepackage{xcolor}
\usepackage{multirow}
\def\BibTeX{{\rm B\kern-.05em{\sc i\kern-.025em b}\kern-.08em
    T\kern-.1667em\lower.7ex\hbox{E}\kern-.125emX}}

\usepackage[linesnumbered,ruled]{algorithm2e}
\SetKwInput{KwInput}{Input}                %
\SetKwInput{KwOutput}{Output}              %

\usepackage{subfig}

\usepackage{cleveref} 
            
\begin{document}
\bstctlcite{IEEEexample:BSTcontrol}

\title{Hyperion: Low-Latency Ultra-HD Video Analytics via Collaborative Vision Transformer Inference\\
}

\author{\IEEEauthorblockN{Linyi Jiang\IEEEauthorrefmark{1}, Yifei Zhu\IEEEauthorrefmark{1},
Hao Yin\IEEEauthorrefmark{2} and
Bo Li\IEEEauthorrefmark{3} \thanks{The research was supported in part by an NSFC grant 62432008, NSFC grant 62302292, RGC RIF grant R6021-20, an RGC TRS grant T43-513/23N-2, RGC CRF grants C7004-22G, C1029-22G and C6015-23G, NSFC/RGC grant CRS\_HKUST601/24 and RGC GRF grants 16207922, 16207423 and 16203824. Corresponding author: Yifei Zhu. }}
\IEEEauthorblockA{\IEEEauthorrefmark{1}Global College, Shanghai Jiao Tong University\\
\IEEEauthorrefmark{2}Beijing National Research Center for Information Science and Technology, Tsinghua University\\
\IEEEauthorrefmark{3}Guangzhou HKUST Fok Ying Tung Research Institute and Department of Computer Science and Engineering, \\Hong Kong University of Science and Technology\\
Email: jiangly01@sjtu.edu.cn, yifei.zhu@sjtu.edu.cn, h-yin@mail.tsinghua.edu.cn,  bli@cse.ust.hk}}
\maketitle

\begin{abstract}
Recent advancements in array-camera videography enable real-time capturing of ultra-high-definition (Ultra-HD) videos, providing rich visual information in a large field of view.
However, promptly processing such data using state-of-the-art transformer-based vision foundation models faces significant computational overhead in on-device computing or transmission overhead in cloud computing.
In this paper, we present Hyperion, the first cloud-device collaborative framework that enables low-latency inference on Ultra-HD vision data using off-the-shelf vision transformers over dynamic networks.
Hyperion addresses the computational and transmission bottleneck of Ultra-HD vision Transformer inference by exploiting the intrinsic property in vision Transformer models.
Specifically, Hyperion integrates a collaboration-aware importance scorer that identifies critical regions at the patch level, a dynamic scheduler that adaptively adjusts patch transmission quality to balance latency and accuracy under dynamic network conditions, and a weighted ensembler that fuses edge and cloud results to improve accuracy.
Experimental results on real-world prototypes and datasets demonstrate that Hyperion enhances frame processing rate by up to  1.61 \(\times\) and improves the accuracy by up to 20.2\% when compared with state-of-the-art baselines under various network environments.
\end{abstract}

\section{Introduction}
\begin{figure}[t]
\centerline{\includegraphics[width=0.475\textwidth]{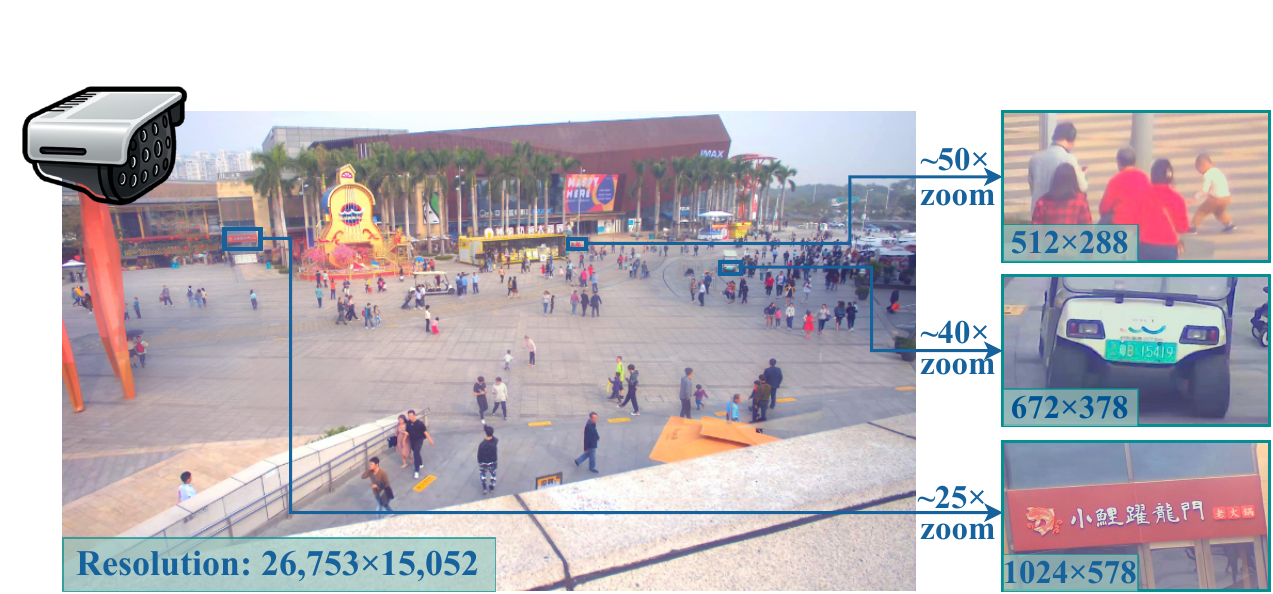} }
\caption{An example of an ultra high definition frame (PANDA dataset \cite{wang2020panda}) captured by an array camera and its details under different zoom ratios.}
\label{fig:Intro}
\end{figure}

Recent developments in array-camera videography techniques have enabled real-time capture of Ultra-HD videos, in 4K, 8K, or even gigapixel quality \cite{chen2022scaling, yuan2017multiscale, wang2020panda}. 
In contrast to traditional cameras that struggle with the tradeoff between field of view and resolution, an array camera, formed by 
a number of micro-cameras, can capture an expansive field of view ($\sim$1$km^2$) at Ultra-HD resolutions (reaching up to 26{,}753×15{,}052 resolution) in a single frame, as illustrated  in Fig.~\ref{fig:Intro}. 
Analyzing such videos is critical for applications that require wide spatial awareness and great detail comprehensiveness(e.g., situation-aware surveillance in public squares), or scenarios where wide deployment of a camera network is physically impractical or costly (e.g., wildlife protection in safaris and coastline monitoring).
For these applications to be effective, Ultra-HD videos need to be processed and analyzed with both low latency and high accuracy to ensure accurate and timely responses.

Vision transformers (ViTs), which are vision foundation models based on transformer architecture, have outperformed traditional model architectures (e.g., CNNs, RNNs), achieving state-of-the-art (SOTA) results across  a broad range of computer vision tasks, including image classification \cite{dosovitskiy2020image}, object detection \cite{li2022exploring}, and semantic segmentation \cite{thisanke2023semantic}.
Their ability to accurately process complex visual inputs makes them a strong enabler for extracting rich insights from Ultra-HD video streams in next-generation video analytics systems.

While Vision Transformers offer promising accuracy, achieving low-latency Ultra-HD visual processing with these models remains challenging due to substantial computational and communication overheads.
On the one hand, as vision transformers are inherently resource-intensive, on-device computing of these ultra-HD videos is often infeasible. Specifically, the self-attention mechanism in ViTs introduces \textit{quadratic computational complexity} and inference latency as the input size increases \cite{tay2022efficient, han2022survey}, making them ill-suited for high-resolution inputs on resource-constrained edge devices. 
For example, Jetson Orin Nano\cite{nvidia} can only serve the full-precision ViTDet-Base\cite{li2022exploring} at 0.46 FPS for standard HD (i.e., 1080×720 resolution) input, rendering Ultra-HD processing infeasible.
On the other hand, offloading computation to remote cloud servers equipped with powerful accelerators offers an alternative but introduces significant communication overhead. The illustrative frame in Fig.~\ref{fig:Intro} has a size of 18 MB, taking 9.83 seconds to transmit over commercial 5G networks \cite{jackson2023ofcom}. 
These combined challenges of compute intensity and bandwidth demand significantly hinder the deployment of ViTs for Ultra-HD video analytics in latency-sensitive, real-world scenarios.

Existing system-level video analytics solutions largely focus on conventional CNN-based models operating at standard or lower resolutions. These systems typically address the communication-accuracy tradeoff through adaptive streaming, region of interest(ROI) extraction, or frame filtering. 
While early efforts such as Elf \cite{zhangElfAccelerateHighresolution2021a} extend support to high-resolution input, their reliance on coarse-grained ROI selection (at the object or even larger block level) leads to substantial latency  degradation when applied to vision transformers on Ultra-HD video.
Parallel efforts at the model level design techniques such as attention approximation\cite{yuanHrformerHighresolutionVision2021} and hierarchical processing\cite{yuMegabytePredictingMillionbyte2023} to improve inference efficiency. However, these solutions are primarily designed for powerful servers and require architecture modification to existing models. As such, they lack compatibility with commercial off-the-shelf ViTs and are unsuitable for deployment on resource-constrained edge devices. 
Furthermore, intense reliance on cloud servers introduces fragility when network connectivity is intermittent or unavailable.  
Therefore, designing an Ultra-HD video analytics system that is \textit{efficient}, \textit{generalizable} to off-the-shelf vision transformers and \textit{deployable} in resource-limited settings remains an open and pressing challenge.

In this paper, we introduce Hyperion, a collaborative cloud-device Ultra-HD video analytics system tailored for ViT inference over dynamic networks. 
\textit{The system builds on the key insight that ViTs inherently partition input images into patches, offering a model-native opportunity for fine-grained system optimization.}  
Leveraging this, Hyperion generalizes the traditional coarse-grained ROI concept into a fine-grained Patch of Interest (POI) abstraction, aligning system optimization with the ViT computation model. It first jointly considers the collaborative cloud-device architecture and model-induced attention score to assess the importance of each patch. Hyperion then dynamically controls patch quality for efficient communication, guided by a lightweight patch-level profiler and an optimal transmission scheduler. Finally, a weighted model ensembler integrates edge and cloud inference results to enhance accuracy further.
These designs enable Hyperion to exploit both ViT model structure and cloud-device computing architecture for low-latency, high-accuracy inference under dynamic and even intermittent networks. 
To the best of our knowledge, it is the first system to realize efficient, low-latency Ultra-HD inference for off-the-shelf vision transformers.
In summary, our contributions are provided as follows:
\begin{itemize}
\item We analyze the computational characteristics of ViTs on Ultra-HD inputs, revealing patch-level opportunities for resource optimization within practical system constraints
(Section \ref{sec:Background and Motivation}). 
\item We design Hyperion, a novel collaborative cloud-device inference system specifically designed to enable low-latency, resource-efficient inference of off-the-shelf ViTs on Ultra-HD video streams (Section \ref{sec:Background and Motivation}). 
\item We propose a collaboration-aware importance scorer to identify patches of interest (Section \ref{sec:Attention-Based Patch-level Scorer}). To adapt to dynamic network conditions, we design a dynamic programming-based scheduler to determine the appropriate quality level for each patch (Section \ref{sec:Profiler}, \ref{sec:Adaptive Transmission Scheduler}).
\item  We implement Hyperion on real-world devices and evaluate it under practical network environments. Hyperion improves frame processing rate by up to 1.61× and accuracy by 20.2\% compared to state-of-the-art baselines (Section \ref{sec:Evaluation}).
\end{itemize}

\section{Background and Motivation Study}\label{sec:Background and Motivation}

\subsection{Transformer-based Vision Model}\label{sec:vision transformer}

\begin{figure}[t]
\centerline{\includegraphics[width=0.475\textwidth]{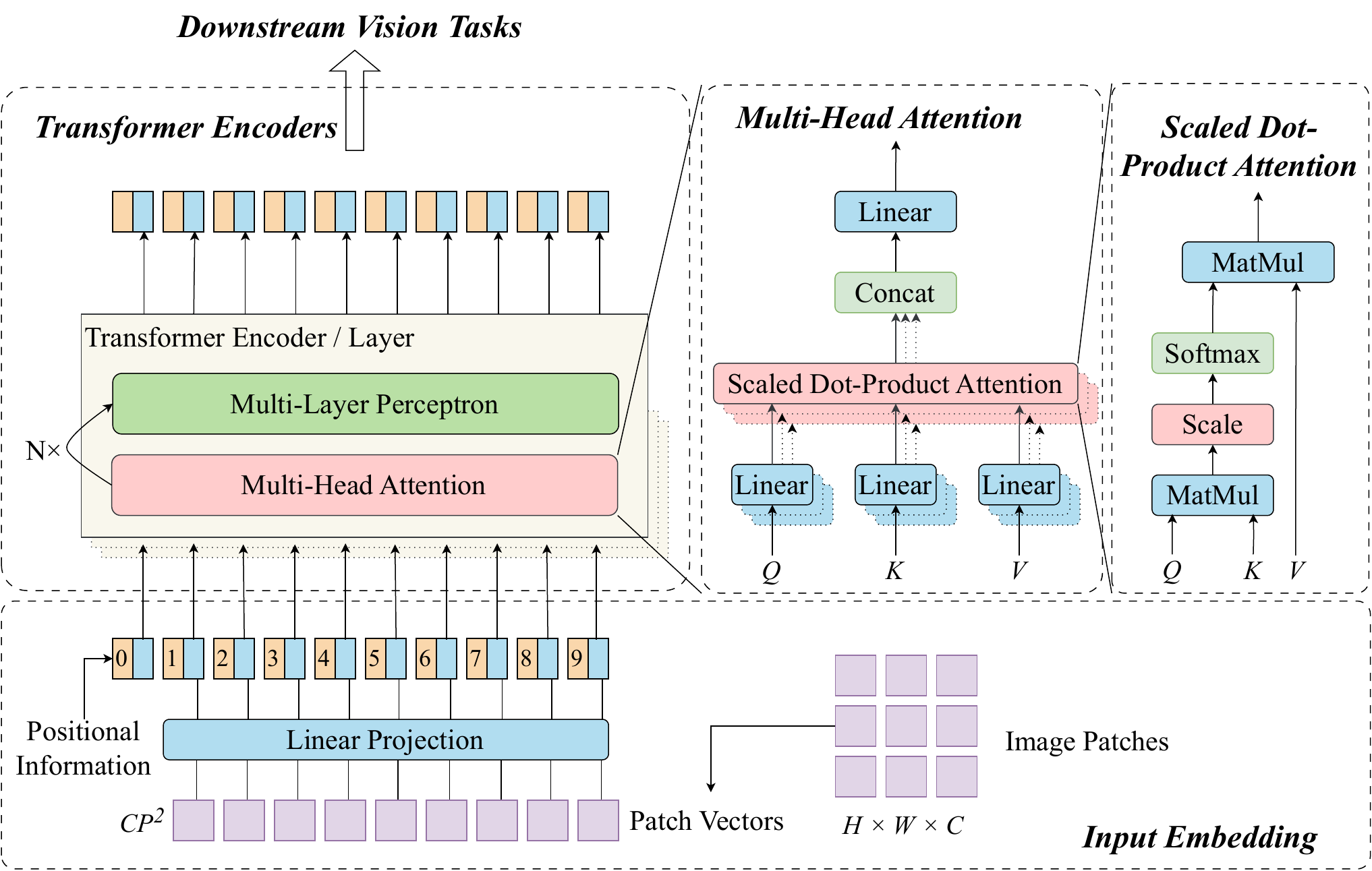} }
\caption{The typical architecture of a vision transformer.}
\label{fig:ViT}
\end{figure}

Transformer-based vision models are composed of a series of transformer encoder layers. As shown in Fig. \ref{fig:ViT},
models first divide the input image, with dimensions $H \times W \times C$ (height, width, channels), into $n = \frac{H \times W}{P^2}$) non-overlapping patches, each of size $P \times P$. These patches are then flattened into vectors of length $CP^2$, and positional information is added to retain spatial context so that we have complete input embedding vectors for the transformer encoders.
The transformer encoder includes a multi-head attention (MHA) module that computes attention scores, followed by a multi-layer perceptron(MLP) to introduce nonlinearity \cite{vaswani2017attention}.
As this process is repeated across $N$ layers, the model effectively encodes complex features from the image and can be further utilized for various downstream vision tasks.

The self-attention mechanism is fundamental to transformer-based vision models, enabling the MHA module to capture the relationships between image patches.
In the MHA module, the input \( X \in \mathbb{R}^{n \times d} \) is first projected into Query ($Q = XW_q$), Key ($K = XW_k$), and Value ($V = XW_v$) representations using learned weight matrices $W_q$, $W_k$, and $W_v$. Each of these representations is then split into multiple heads to perform the attention function in parallel. 
For a given head $h$ and layer $l$, the attention weights $A$ are computed as:
\begin{equation}
\mathbf{A}^{(h, l)}\left(\mathrm{p}_i, \mathrm{p}_j\right) = \operatorname{softmax} \left(\frac{\mathbf{Q} \mathbf{K}^T}{\sqrt{d}} \right)_{(i, j)} \in \mathbb{R}^{n \times n}
\end{equation}
where $i$, $j$ index patches in the image, and $d$ is the dimensionality of $K$ for each attention head. The attention weights quantify the relevance of patch $p_j$ to patch $p_i$ within a specific attention head and layer. 
The attention weights are then multiplied by $V$, producing the output for a single attention head. Finally, the outputs of attention heads are concatenated, forming the final output of the MHA module. %
Notably, the computational bottleneck lies in the matrix multiplications within the MHA module: computing \( QK^\top \) and the subsequent multiplication with \( V \), both of which have a complexity of \( \mathcal{O}(n^2) \). Thus, the overall self-attention mechanism scales quadratically with the number of patches \( n \), making it prohibitively expensive for high-resolution inputs.

\subsection{Motivation Study}\label{sec:Motivation Study}
Based on this understanding of vision transformers, we begin by analyzing Ultra-HD inputs from object-level to patch-level behavior in the context of object detection, the dominant task in existing Ultra-HD video analytics applications.

{\bf Vision dataset and model:}
For the dataset, we downsample surveillance video footage from the  de facto PANDA dataset\cite{wang2020panda} to Ultra-HD  resolution (3840×2160). The scenes depict large-area environments such as public squares and open playgrounds. 
For the inference model, we employ the ViTDet-Base\cite{li2022exploring}, representing a typical architecture in the family of transformer-based vision models. 
\begin{figure}[t]
	\centering
	\subfloat[\label{a} Input image example.]{
		\includegraphics[width=0.212\textwidth]{COCOExample.jpg}}
	\subfloat[\label{b} Patch importance heatmap.]{
		\includegraphics[width=0.258\textwidth]{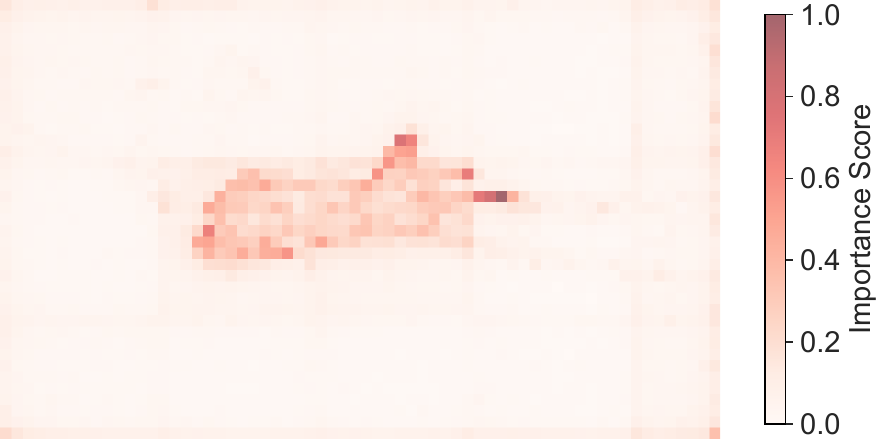} }
	\caption{Visualization of importance scores generated during the inference of vision transformers.}
	\label{fig:MotivationSetup} 
\end{figure}

{\bf Patch-level importance score:}
As mentioned in Section~\ref{sec:vision transformer}, the attention weights $A$ represent the level of attention each image patch allocates to others. As a common metric for importance analysis in vision transformers~\cite{fayyaz2022adaptive, wang2024zero}, we compute a patch-level importance score by averaging attention weights across all heads and layers in the output feature layers.
A higher score indicates that a patch is more influential within the context of the image. 
To illustrate, Figure \ref{fig:MotivationSetup} shows an example image processed by ViTDet-Base, where object regions (e.g., the airplane) exhibit significantly higher importance scores than background areas, reflecting their prominence in the attention mechanism. We leverage this attention-based scoring to study the behavioral characteristics of Ultra-HD ViT inference and inform system-level optimization.
For interpretability,  we apply the Jenks Natural Breaks Clustering algorithm\cite{jenks1967data} to partition the continuous importance scores into three discrete classes, with higher class levels indicating greater significance. 

\subsection{Observations and Implications} \label{sec:Observations} 

\textbf{Observation 1: Inference difficulty varies across object sizes  in Ultra-HD frames.}
We begin our analysis at the object level, focusing on how object size impacts inference difficulty  for ViTs in Ultra-HD video. Ultra-HD frames often contain objects of widely varying sizes, and prior work has shown that smaller objects are significantly harder to detect than larger ones for conventional models\cite{wang2020panda}.
To validate this on ViTs, we evaluate ViTDet-Base across different object sizes: small objects (width and height $\leq$32 pixels) and large objects (width and height $\geq$96 pixels).
ViTDet-Base with 111M parameters achieves only 1.97\% mAP on Small objects (i.e., APS), while reaching 12.19\% mAP on Large objects (i.e., APL), confirming the increased difficulty of detecting small objects even using ViTs. 
We further assess ViTDet-Large with 331M parameters under the same conditions. While it improves APS by approximately 23\%, it yields less than a 4\% gain in APL. This highlights that larger models benefit small-object detection but offer minimal improvements for large-object regions at significantly higher computational cost.
\textit{
This observation essentially calls for different treatments of small and large objects to strike a balance between model accuracy and computational cost. A collaborative inference architecture where a lightweight model at the edge is used to process regions of large objects, while offloading only small object regions to a more powerful cloud is motivated. 
}

\begin{figure}[t]
	\centering
	\subfloat[\label{a} Proportion of patch importance class within frame.]{
		\includegraphics[width=0.24\textwidth]{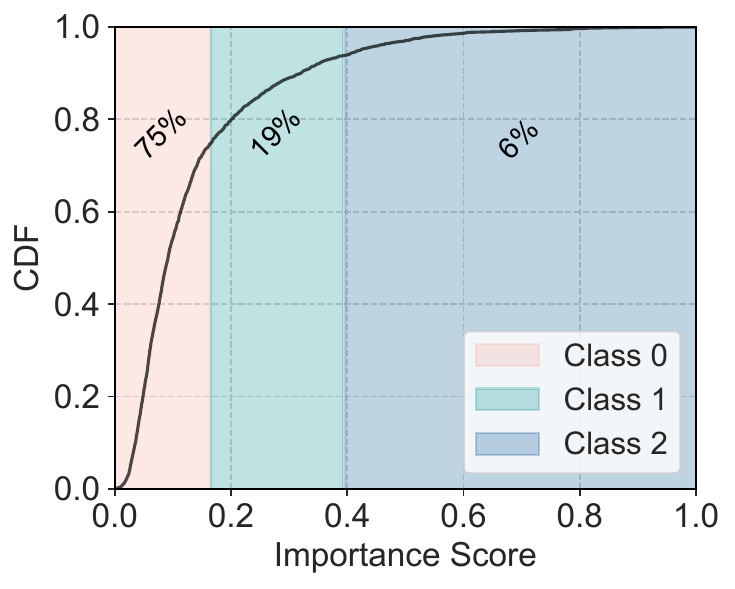}}
	\subfloat[\label{b} Bandwidth Consumption and\\Accuracy vs. Downsampling Rate. ]{
		\includegraphics[width=0.24\textwidth]{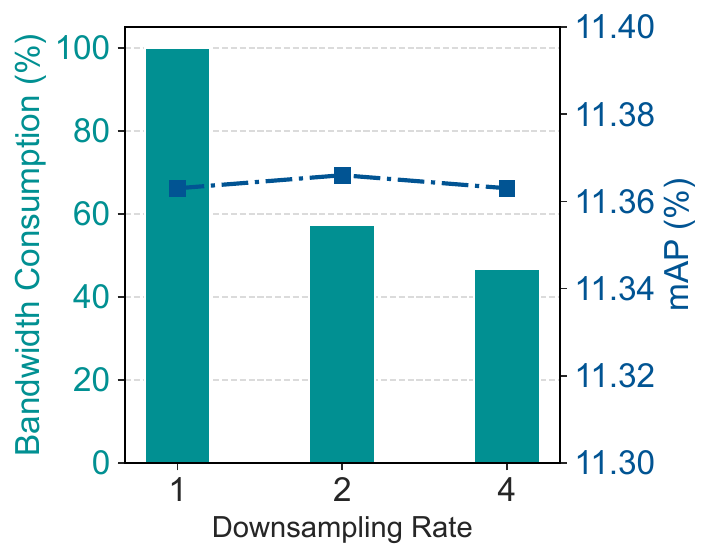} }
	\caption{The distribution of attention weights within the frame is biased.}
	\label{fig:Obs1} 
\end{figure}
{\bf Observation 2: The distribution of importance scores within the Ultra-HD frame is biased.}
Next, we shift our focus from object-level characteristics to patch-level, investigating the spatial distribution of importance scores within the frame. Figure \ref{fig:Obs1} (a) illustrates the cumulative distribution function (CDF) of the importance scores for each patch class in the frames of scene \textit{University Playground} in the dataset. The results indicate a biased distribution of attention weights, with 74.74\% of the patches falling into Class 0. This suggests that a large portion of the frame contributes little to the task. By examining these patches, we find that patches in Class 0 primarily correspond to background content. 
This bias can be attributed to the characteristics of Ultra-HD data, which often capture wide fields of view that contain significant amounts of background information.
To investigate the potential for resource optimization, we perform downsampling on patches in Class 0 and assess the impact on average bandwidth consumption and detection accuracy across all scenes. Note that in our pilot study, the bandwidth consumption calculation does not consider any image compression algorithms. As shown in Figure \ref{fig:Obs1} (b), compared to transmitting the entire frame (i.e., downsampling rate = 1), downsampling the Class 0 patches to a rate of 4 can reduce bandwidth consumption by 53.33\%, without any drop in accuracy.
\textit{This observation suggests that ultra-HD frames contain a disproportionate amount of insignificant patches regarded by the vision transformer. 
Transmitting these less important patches at reduced quality can significantly reduce bandwidth consumption while maintaining the high accuracy achieved by vision transformers.
}

\begin{figure}[t]
	\centering
	\subfloat[\label{a} Proportion of patch importance class within object.]{
\includegraphics[width=0.235\textwidth]{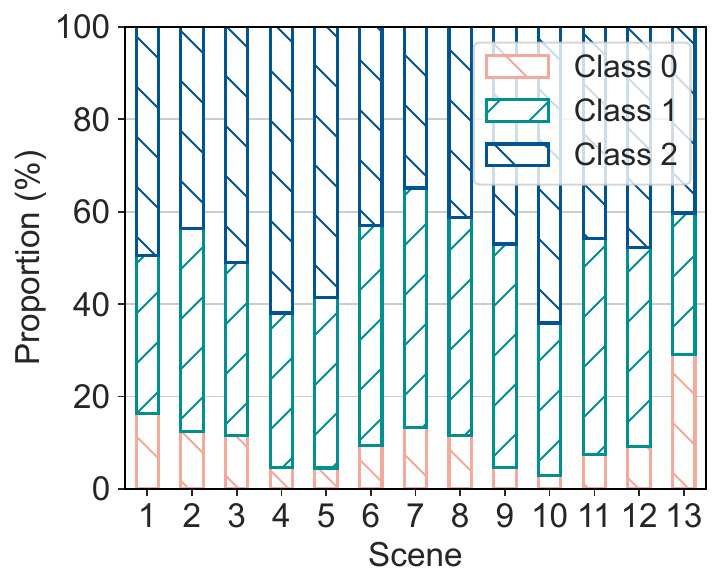}}
	\subfloat[\label{b} Bandwidth Consumption and\\Accuracy vs Downsampling Rate. ]{
\includegraphics[width=0.235\textwidth]{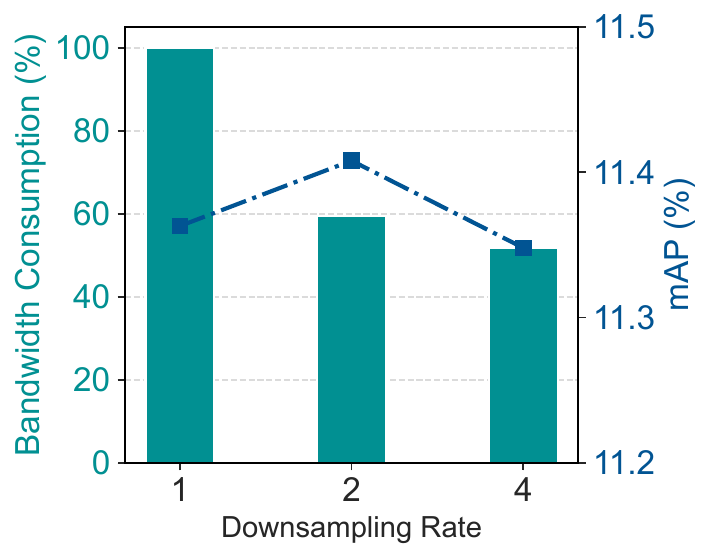} }
	\caption{The vision transformer backbone is not sensitive to the completeness of the object area.}
	\label{fig:Obs2} 
\end{figure}

\textbf{Observation 3: Vision transformers are insensitive to object completeness.}
We conclude our analysis by examining how the importance scores are distributed within object regions. Using ground-truth bounding box annotations, we compute the proportion of each patch importance class contained within object areas. 
As shown in Figure \ref{fig:Obs2},  the most influential regions (Class 2) account for less than half of the object area, while 9.71\% of the object region still belongs to Class 0, the lowest important class. 
This indicates that not all regions within an object contribute equally to the model's predictions.
In contrast to prior approaches that treat all object patches uniformly, our results show that Vision Transformers are relatively insensitive to the completeness of object regions. Uniformly transmitting all patches within a detected object leads to redundant data transfer and inflated bandwidth usage. To evaluate the benefits of finer-grained control, we simulate patch-level downsampling within object areas. Specifically, we apply 2x and 4x downsampling to Class 1 patches while keeping Class 0 patches fixed at 4x downsampling. As shown in Figure \ref{fig:Obs2} (b), compared to object-level ROI transmission (i.e., downsampling rate = 1), this POI strategy reduces bandwidth by 40.43\%, with no loss in detection accuracy when the downsampling rate is two.
\textit{This observation highlights patch-level importance-aware adaptation as a more efficient alternative to object-level approaches for communication-efficient Ultra-HD ViT inference.
}
\begin{figure}[t]
\centerline{\includegraphics[width=0.5\textwidth]{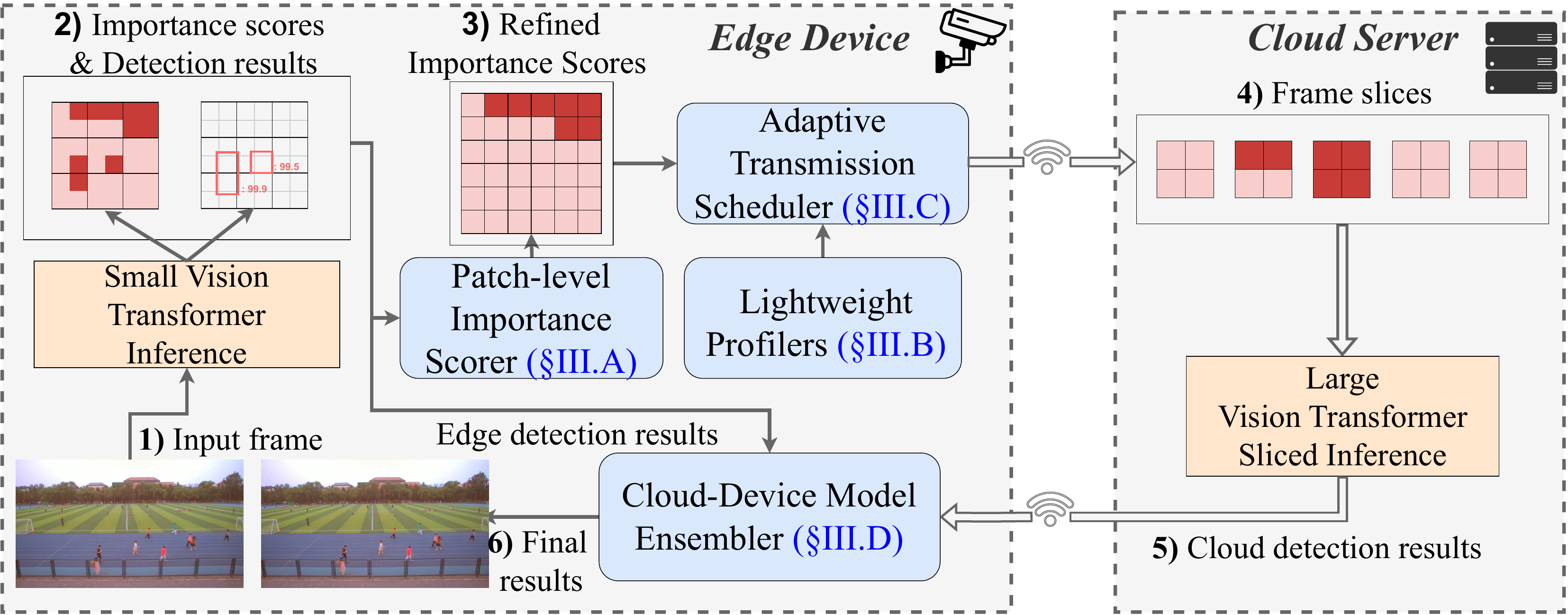} }
\caption{System overview of Hyperion.}
\label{fig:SystemOverview}
\end{figure}

\section{System Design}\label{sec:System Design}
Our motivation study reveals the opportunities for resource savings in Ultra-HD inference systems using vision transformers, ranging from within-frame to within-object optimizations at the patch level. However, developing a resource-efficient system still faces several critical challenges. 
First, assigning an importance score to each patch is non-trivial. 
Directly using attention-derived scores (as mentioned in the motivation study) overlooks the collaborative inference context and may send those easy-to-analyze regions to the cloud, incurring unnecessary transmissions.  Another key challenge is making appropriate transmission decisions for each patch, especially under dynamic network conditions where latency constraints must be met. 
Furthermore, to efficiently adapt transmission decisions, it is crucial to estimate the transmission latency and the accuracy impact of different strategies with low overhead.

To address these challenges, we propose Hyperion, a collaborative inference framework that leverages both edge devices and cloud servers to achieve low-latency and high-accuracy Ultra-HD vision transformer inference.
At the edge device, a lightweight vision transformer identifies high-priority patches and provides baseline predictions when network conditions degrade, ensuring system responsiveness and reliability. Concurrently, the cloud runs a more powerful model to process the selected patches, refining results and improving overall inference accuracy.

Fig. \ref{fig:SystemOverview} provides an overview of the Hyperion system architecture. 
Specifically, the \textbf{collaboration-aware patch-level importance scorer} (\S \ref{sec:Attention-Based Patch-level Scorer}) first calculates and refines the importance scores of each patch within the frame, considering both model-specific attention scores and the collaborative inference architecture.  We deploy \textbf{lightweight profilers} (\S \ref{sec:Profiler}) to predict frame size and accuracy under various conditions. 
Based on importance scores and profilers, the \textbf{dynamic transmission scheduler} (\S \ref{sec:Adaptive Transmission Scheduler}) selects the appropriate quality level for each patch, accounting for latency constraints and dynamic network conditions. The processed frame is then transmitted to the cloud server, where it is further analyzed using a larger vision transformer. Finally, the \textbf{cloud-device model ensembler} (\S \ref{sec:ModelEnsembler}) at the edge integrates results from both models to produce the final output.

\subsection{Collaboration-aware Patch-level Importance Scorer}\label{sec:Attention-Based Patch-level Scorer}
Existing importance scoring in vision transformers is model-centric and unaware of collaborative device-cloud inference contexts.
To address this, we propose a collaboration-aware patch-level scorer that refines attention-based scores by considering the device-cloud performance gap.
This enables the system to further reduce transmission overhead by limiting high-quality transmission to only essential patches.

As discussed in Section \ref{sec:Motivation Study}, considering the device-cloud performance gap, patches within large objects that can be adequately analyzed by the small vision transformer on the edge do not require high-quality transmission, while those patches within small objects should be sent to the cloud for further analysis to preserve accuracy. This distinction is supported by our prior analysis based on ground truth object sizes. However, since true object sizes are unknown at runtime, we use the confidence scores from the edge model’s inference as a practical proxy to identify hard-to-analyze regions, which often correspond to small objects. So to identify these patches, we use a combination of detection results (i.e., the bounding boxes of detected objects and their associated confidence scores) and patch importance scores. 
These data are generated directly during the inference process of the small vision transformer, without requiring any additional computations.

We first define the patch importance scores derived from the attention mechanism of the vision transformer. To aggregate the importance of patch $p_i$ across layers, heads, and tokens, we compute the importance score as follows:
\begin{equation}
ImpScore\left(\mathrm{p}_i\right) = \frac{1}{L N_h n} \sum_{l=1}^{L} \sum_{h=1}^{N_h} \sum_{j=1}^n \mathbf{A}^{(h, l)}\left(\mathrm{p}_i, \mathrm{p}_j\right)
\end{equation}
where $L$ is the total number of output feature layers, $N_h$ is the number of attention heads, and $n$ is the number of patches. 

We next group patches into $K$ distinct classes using a mini-batch version of the Jenks Natural Breaks algorithm\cite{jenks1967data},  assigning each patch 
$p_i$ a class label $c_p \in C = \{0, 1, ..., K-1\}$.
This class-based approach reduces the complexity of the optimization process (Section \ref{sec:Adaptive Transmission Scheduler}), making it more computationally efficient. 
We further refine patch priorities by applying two thresholds: an importance threshold $T_i$ to determine whether a region is important from a ViT's perspective, and a detection confidence threshold $T_c$ to determine whether a region within the corresponding bounding box is easily analyzable for edge models.
Patches that exceed both thresholds are considered easily analyzable by the edge model. These patches are reassigned to the lowest importance class (i.e., $c_p = 0$), effectively lowering their transmission priority.
Notably, rather than directly pruning patches with low importance class, we adjust their transmission quality which preserves essential background context and inter-object semantics \cite{zhangElfAccelerateHighresolution2021a, cao2023edge} while still achieving bandwidth reduction. It aligns with our goal of minimizing transmission overhead without sacrificing accuracy.

\subsection{Lightweight Profilers}  \label{sec:Profiler}
\begin{figure}[t]
    \centering
    \subfloat[Compression ratio vs. weighted average compression quality.]{
        \includegraphics[width=0.235\textwidth]{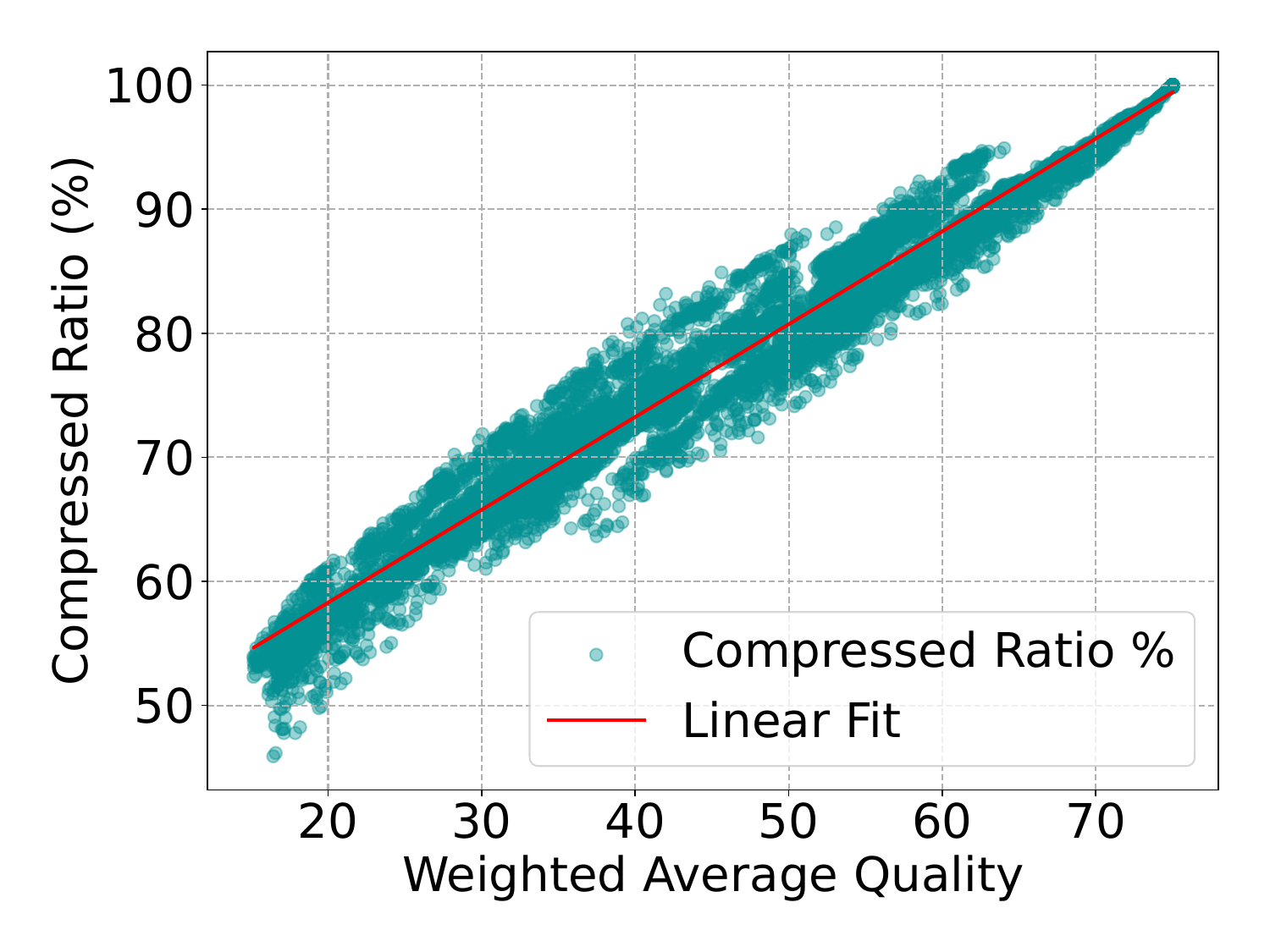}
    }
    \subfloat[Accuracy loss vs. quality across different classes.]{
        \includegraphics[width=0.235\textwidth]{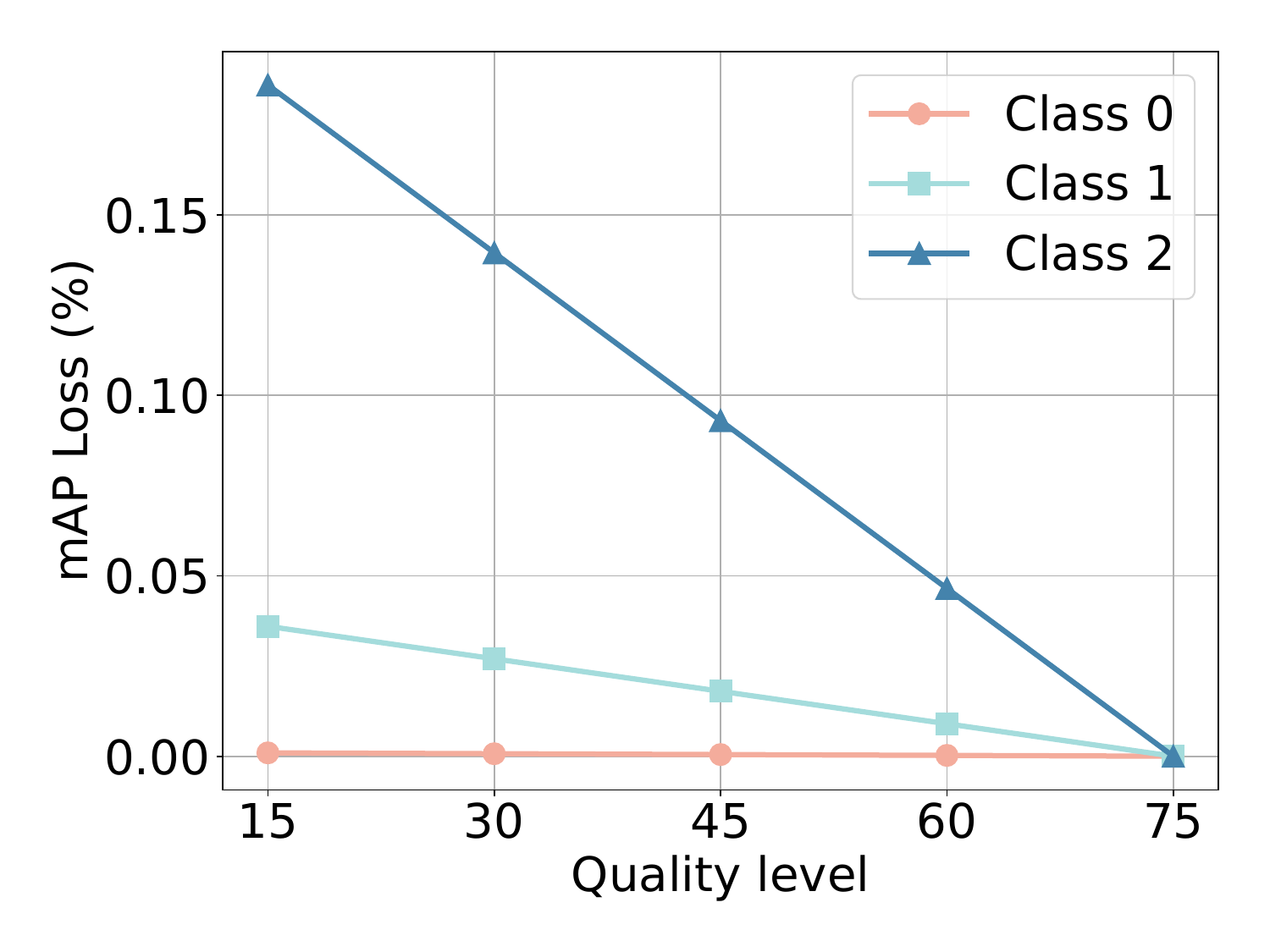}
    }
    \caption{Profiling the impact of quality settings on transmission size and accuracy.}
    \label{fig:profiler}
\end{figure}

Selecting the suitable transmission configuration for low-latency and high-accuracy inference in dynamic networks requires considering the expected transmission size and accuracy. 
The transmission size corresponds to the compressed frame size after applying a given quality level to each patch. 
During scheduling, estimating latency for different quality values would require repeated compression operations, resulting in substantial computational overhead. 
Moreover, to ensure high overall accuracy, it is essential to estimate the accuracy performance at different quality values. 
To address these challenges, we propose two offline profilers: a \textit{Size Profiler} to estimate compressed frame size and an \textit{Accuracy Profiler} to predict accuracy under different quality configurations.

We experiment by applying all possible combinations of available quality levels to different classes of patches and analyzing the resulting compression ratios and accuracy across various frames. 
In this experiment, the number of classes $K$ is set to 3, and the available JPEG quality values $Q$ are set to $\{15, 30, 45, 60, 75\}$. Each patch class $c_p$ is assigned a quality level $q_k \in Q$, and the selected quality levels across all classes are denoted as $Q_c = \{q_0, q_1, \dots, q_{K-1}\}$.
The experiments use the ViTDet-base and the PANDA-Image dataset. 

\textbf{Size Profiler.} 
Fig. \ref{fig:profiler}(a) shows the relationship between compression ratio (i.e., the ratio of the compressed frame size to the original frame size) and weighted average compression quality (i.e., the sum of patch-wise quality values weighted by their proportions in each frame), which represents the overall quality of the entire frame. Our observations reveal a strong positive linear correlation between these two variables across diverse scenes. The correlation coefficient exceeds 0.98, with a P-value near 0, indicating statistical significance. Therefore, we model the compressed frame size $S (Q_c)$ as: 
\begin{equation}
    S (Q_c) = \underbrace{\left( \alpha \sum_{c=0}^{K} w_c q_c + \alpha_S \right)}_{\text{\small Compression Ratio}} \times S_O
\end{equation}
where $S_O$ denotes the original frame size, $w_c$ is the proportion of patches belonging to class $c$, $q_c$ is the corresponding quality factor, and $\alpha$ and $\alpha_S$ are regression coefficients.

\textbf{Accuracy Profiler.} 
We also find a strong multivariate linear correlation between accuracy and patch quality levels, with a correlation coefficient over 0.84 and a $p$-value of $5.07 \times 10^{-7}$.
Therefore, we can model accuracy $A (Q_c)$ as:
\begin{equation}
    A (Q_c) = \sum_{c=0}^{K} \beta_c q_c + \beta_A.
\end{equation}
where $\beta_k$ and $\beta_A$ are regression coefficients. 

Fig. \ref{fig:profiler}(b) further illustrates how accuracy degrades across patch classes under reduced quality. Notably, accuracy degradation varies significantly across classes. Low-importance classes (e.g., Class 0) are highly robust against quality reduction, with almost no accuracy loss. As the importance of the patch increases, there is a significant decrease in accuracy with lower quality values.

The \textit{size profiler} and \textit{accuracy profiler} take the quality levels selected for each patch class in the current frame as input and output the predicted compressed frame size and accuracy.
We calculate the model coefficients using linear regression, which aligns with real-world behaviors and reduces computational overhead, simplifying the generation of prediction models.

\subsection{Adaptive Transmission Scheduler}\label{sec:Adaptive Transmission Scheduler}

Next, we design the adaptive transmission scheduler to optimize the transmission process of patches, aiming to maximize inference accuracy while ensuring that latency remains within the constraint for processing each frame. This requires selecting appropriate quality values for each patch under dynamic network conditions.

Due to the similar behavior of patches within a class, we simplify the problem by determining the optimal quality value at a class level. Specifically, we seek the optimal set \( Q_c^* = \{q_c^* \mid c \in C, q_c^* \in Q\} \) that maximizes the overall accuracy \( A \) while ensuring that the total latency remains within the constraint \( L \). The transmission scheduling problem can be formulated as the following optimization problem:

\begin{equation}
\begin{aligned}
\max \quad & \sum_{c=0}^{K-1} \beta_c q_c + \beta_A \\
\text{s.t.} \quad
& \begin{cases}
\frac{{\left( \alpha \sum_{c=0}^{K-1} w_c q_c + \alpha_S \right)} \times S_O}{B} + L_\text{d} + L_\text{c} \leq L, \\
q_c \in Q, \quad \forall c \in C.
\end{cases}
\end{aligned}
\end{equation}
where $L_\text{d}$ is device processing latency, $L_\text{c}$ is cloud processing latency, $B$ is the predicted network bandwidth.
This problem can be modelled as a variant of the Multiple-Choice Knapsack Problem (MCKP)\cite{pisinger1995minimal}, for which we design a dynamic programming-based algorithm in Algorithm \ref{alg:scheduler} to solve it optimally. 
The scheduler first converts the latency constraint into the maximum frame size using the estimated current network bandwidth and the device and cloud processing latencies (line 2). We estimate the bandwidth based on the harmonic mean of the observed throughput \cite{jiang2012improving}. For cloud processing latency, we profile the inference latency on target devices offline. It then initializes a dynamic programming (DP) table to track maximum accuracy values and a path matrix to record quality selections (line 3).
The core procedure iterates over each patch class and possible accumulated frame sizes (lines 4-6). Each valid state evaluates all quality options by calculating their size contributions (lines 7-9). The DP table is updated when a higher accuracy is found for a given frame size
(lines 10-13).
Finally, the algorithm returns the quality configuration corresponding to the maximum accuracy state for all classes (lines 15-18). If no solution satisfies the latency constraint, the system falls back to using device-only inference results.

The time complexity of the scheduler is \( O(K \times |Q| \times S_{max}) \), where $S_{max}$ is the maximum frame size defined in algorithm \ref{alg:scheduler}. Empirical evaluations show that the scheduling overhead is around 2.5 ms, making it suitable for real-time applications.

\begin{algorithm}[t]
\small
\SetAlgoNoEnd  %
\DontPrintSemicolon
\SetAlgoLined
\KwIn{
  \\Class set $C = \{1,\dots,K\}$ with weights $w_c$; 
  Quality values $Q$; Original size $S_O$; 
  Bandwidth $B$; Latency constraint $L$; 
  Device latency $L_d$; Cloud latency $L_c$; 
  Profiled coefficients $\alpha$, $\alpha_S$, $\beta_c$, $\beta_A$
}
\KwOut{
  \\Optimal quality selection $\{q_c^*\}_{c=1}^K$
}
{\bf Procedure:}\\
\text{Compute Maximum Frame Size: }
$S_{\text{max}} \gets \left\lfloor \frac{B(L - L_d - L_c)}{\alpha S_O} - \frac{\alpha_S}{\alpha} \right\rfloor$\;
\BlankLine
\text{Initialize: } 
$DP \gets [-\infty]^{(S_{\max}+1)}$; $DP[0] \gets \beta_A$;
$Path \gets [\emptyset]^{(K+1)(S_{\max}+1)}$\;
\For{$c \gets 1$ \KwTo $K$}{
  $DP_{\text{new}} \gets [-\infty]^{(S_{\max}+1)}$\;
  \For{$s \gets 0$ \KwTo $S_{\text{max}}$}{
    \If{$DP[s] \neq -\infty$}{
      \ForEach{$q \in Q$}{
        $s' \gets s + \lfloor \alpha w_c q \rfloor$\;
        \If{$s' \leq S_{\text{max}}$ \textbf{and} $DP[s] + \beta_c q > DP_{\text{new}}[s']$}{
        $DP_{\text{new}}[s'] \gets DP[s] + \beta_c q$\;
        $Path[c][s'] \gets (q, s)$
      }
  }
    }
  }
    $DP \gets DP_{\text{new}}$\;
}
$s^* \gets \mathop{\arg\max}\limits_{0 \leq s \leq S_{\text{max}}} DP[s]$\;
\For{$c \gets K$ \KwTo $1$}{
  $q_c^* \gets Path[c][s^*].q$\;
  $s^* \gets Path[c][s^*].s$\;
}
\Return $\{q_c^*\}_{c=1}^K$\;
\caption{Dynamic programming-based transmission scheduler}
\label{alg:scheduler}
\end{algorithm}

\subsection{Cloud-Device Model Ensembler}\label{sec:ModelEnsembler}
In our cloud-device collaborative architecture, we further design a  cloud-device model ensembler to integrate the detection results from the cloud with the edge-side preliminary analysis, rather than solely relying on cloud output. 
This fusion strategy is motivated by three primary reasons. 
First, previous research\cite{krogh1994neural, maclin1997empirical, wang2023tabi} has demonstrated that combining outputs from multiple models can reduce variance and bias, resulting in improved inference accuracy. Even ensembling a small model with a large model can yield performance gains. Second, the edge-side small model and the cloud-side large model process inputs at different scales. Integrating these outputs allows the system to utilize both fine-grained local information and broader contextual insights\cite{cai2023efficientvit, havtorn2023msvit}. Third, the data transmitted to the cloud are not the original image patches but versions with adjusted transmission quality. This compression may result in information loss. By combining the outputs of both models, the system can compensate for this potential loss.

Specifically, the detection result from the edge model is represented as 
\([E_{x_1}, E_{y_1}, E_{x_2}, E_{y_2}, E_{\text{conf}}]\), while the detection result from the cloud model is represented as 
\([C_{x_1}, C_{y_1}, C_{x_2}, C_{y_2}, C_{\text{conf}}]\). Here, \((x_1, y_1)\) and \((x_2, y_2)\) denote the top-left and bottom-right coordinates of the bounding box, respectively, and \(\text{conf}\) represents the confidence score of the detection. The ensembled result is denoted as \([F_{x_1}, F_{y_1}, F_{x_2}, F_{y_2}, F_{\text{conf}}]\).
The weights assigned to the edge and cloud models are defined as \(w_e = E_{\text{conf}}\) and \(w_c = C_{\text{conf}}\). Using these weights, the ensembled detection results are computed as follows:
\begin{equation}
\begin{aligned}
F_{x_1} &= \frac{E_{x_1} \cdot w_e + C_{x_1} \cdot w_c}{w_e + w_c}, \quad
F_{y_1} = \frac{E_{y_1} \cdot w_e + C_{y_1} \cdot w_c}{w_e + w_c}, \\
F_{x_2} &= \frac{E_{x_2} \cdot w_e + C_{x_2} \cdot w_c}{w_e + w_c}, \quad
F_{y_2} = \frac{E_{y_2} \cdot w_e + C_{y_2} \cdot w_c}{w_e + w_c}, \\
F_{\text{conf}} &= \frac{w_e + w_c}{2}
\end{aligned}
\end{equation}

After generating the fused results for individual partitions, the overlapping partitions prediction results are merged 
using Non-Maximum Suppression (NMS)\cite{neubeck2006efficient}.

\section{Evaluation}\label{sec:Evaluation}
Our evaluation aims to answer four main questions: \textbf{Q1:} Does Hyperion achieve low latency while maintaining accuracy under various dynamic network conditions? \textbf{Q2:} How do network conditions and latency constraints impact the performance of Hyperion? \textbf{Q3:} How do different components of Hyperion contribute to its overall performance? 
\textbf{Q4:} How much overhead does Hyperion introduce? 

We answer the first three questions through simulation experiments on real-world dynamic network traces, and the last one through an overhead analysis using a real-world deployment. 
Before presenting results, we first detail the real-world deployment of our prototype and experimental setup.

\subsection{Real-World Deployment}
We deploy the prototype of Hyperion on a commercial device-cloud setup, consisting of an NVIDIA Jetson Orin Nano \cite{nvidia} as the edge device and Aliyun ECS instance \cite{elastic} as the cloud server. 
The Jetson Orin Nano, a widely adopted edge computing platform, is equipped with an Ampere GPU with 1024 CUDA cores and 32 Tensor cores, an ARM Cortex-A78AE CPU, and 8GB LPDDR5 DRAM. 
It uses an AW-CB375NF wireless network
module for WiFi communication or a Fibocom FM160-EAU module for cellular mobile communication to transmit data
to the server. The cloud server is equipped with 4$\times$Intel Xeon Platinum 8163 CPUs (2.50GHz), 32GB RAM, and Tesla V100 SXM2 GPU with 5120 CUDA cores and 640 Tensor cores, representing a typical cloud server. Both the edge device and the cloud server operate on Ubuntu 20.04 LTS OS and utilize PyTorch v2.0.0.

\subsection{Experiment Setup} \label{sec:Experiment Setup}
{\bf Video datasets.} 
We evaluate Hyperion on the PANDA video dataset \cite{wang2020panda}, the only available public benchmark for large-scale gigapixel video analytics. It is composed of twenty-one real-world scenes, including street, crossroad, public squares, etc. The dataset contains 75,600 frames with annotations for object detection. Considering the common edge device capability, we downsample the frames to standard Ultra-HD (3840×2160) resolution for our evaluation. 

{\bf Network trace datasets.} To assess the effectiveness of Hyperion under real-world network conditions, we utilize the 5G mmWave uplink performance dataset \cite{ghoshal2022indeptha} for the simulation experiment. The dataset includes three measurement scenarios: Static, Walking, and Driving, covering 5G and 4G LTE networks. We utilize this dataset for its dynamic network conditions, including high levels of fluctuations due to challenging scenarios such as blockage and mobility.

{\bf Parameter settings.}
For the importance scorer, we set $K = 3$, $T_c = 0.90$, and $T_i$ as the class boundary between class 0 and class 1.
For object detection, we use ViTDet \cite{li2022exploring} with a ViT-Small backbone on the device and a ViT-Base backbone on the cloud.
Cloud-side inference uses SAHI \cite{akyon2022slicing} with a 1024 slice size, 0.25 overlap ratio, 0.25 merge IoU threshold, and NMS for merging. 
The latency constraint is set to 400 ms per frame, except when evaluating its sensitivity. 

{\bf Baselines.} We compare Hyperion with the following baselines:  
1) \textit{Remix} \cite{jiangFlexibleHighresolutionObject2021a}: This method dynamically selects models (i.e., EfficientDet and its variants) for different regions within a frame to balance accuracy and efficiency under local computation latency constraints. It represents the SOTA for Ultra-HD on-device processing.  
2) \textit{Elf} \cite{zhangElfAccelerateHighresolution2021a}: This method predicts bounding box shifts across frames using an attention-based LSTM and determines which regions should be offloaded for inference on edge servers. It represents the SOTA for high-resolution offloading processing. The cloud-side model is a traditional Mask R-CNN model, as used in the official implementation of Elf.
3) \textit{Elf-V}: This variant uses the same offloading strategy as ELF but replaces the cloud-side detector with the ViTDet with a ViT-Base backbone and SAHI inference as used in Hyperion. It represents the performance of existing SOTA offloading-based processing methods when applied to vision transformers.

{\bf Evaluation Metrics.} We evaluate the performance of Hyperion and the baselines using the following metrics:  
1) \textit{Inference Accuracy.} Measured by AP50 \cite{coco2020}, which computes the Average Precision (AP) at an IoU threshold of 0.5. Notably, if an inference does not meet the latency constraint, the most recent valid detection result from previous frames is used instead.  
2) \textit{Frame Processing Rate.} The number of frames processed per second (FPS).  
3) \textit{Offload Data.} The size of the transmitted data in megabytes.
4) \textit{Latency Deviation Rate}. It is the percentage difference between measured latency and required latency($\max\left(0, \frac{{\text{Latency}_{\text{measured}} - \text{Latency}_{\text{requirement}}}}{{\text{Latency}_{\text{requirement}}}}\right)$). 

\begin{figure}[t]
    \centering
    \subfloat[Performance under 4G network conditions.]{
        \includegraphics[width=0.45\textwidth]{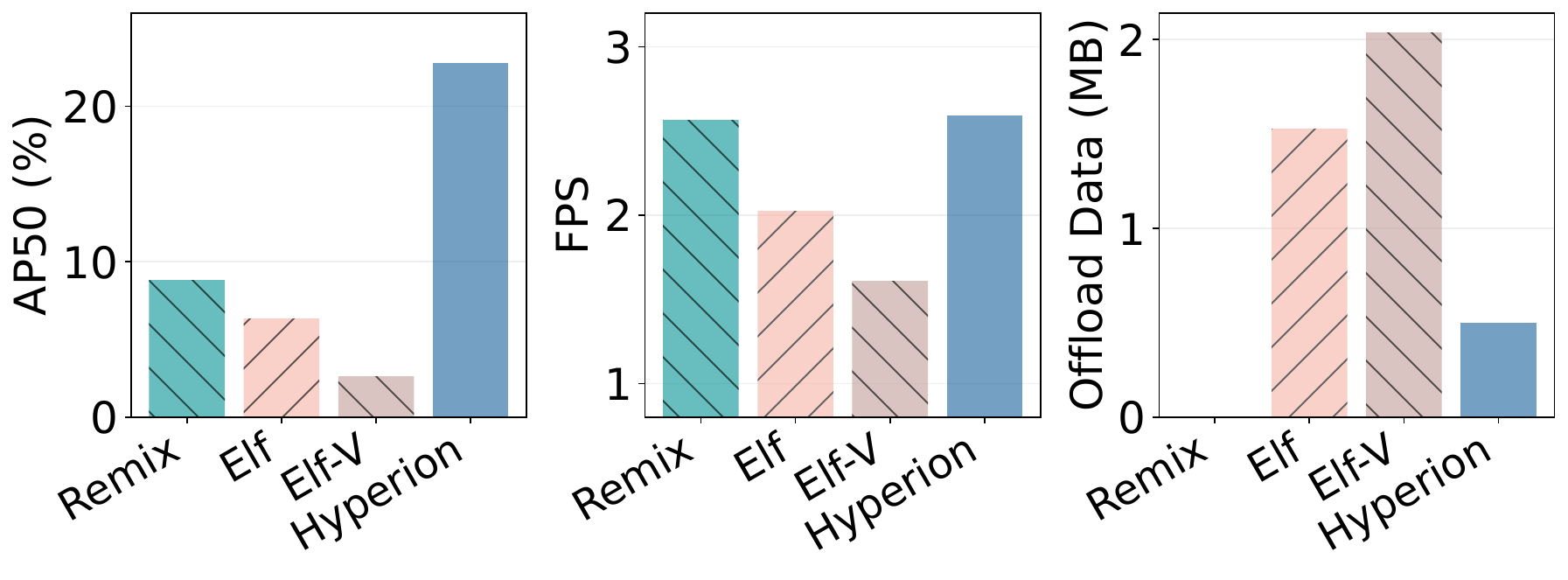}
    }\\
    \subfloat[Performance under 5G network conditions.]{
        \includegraphics[width=0.45\textwidth]{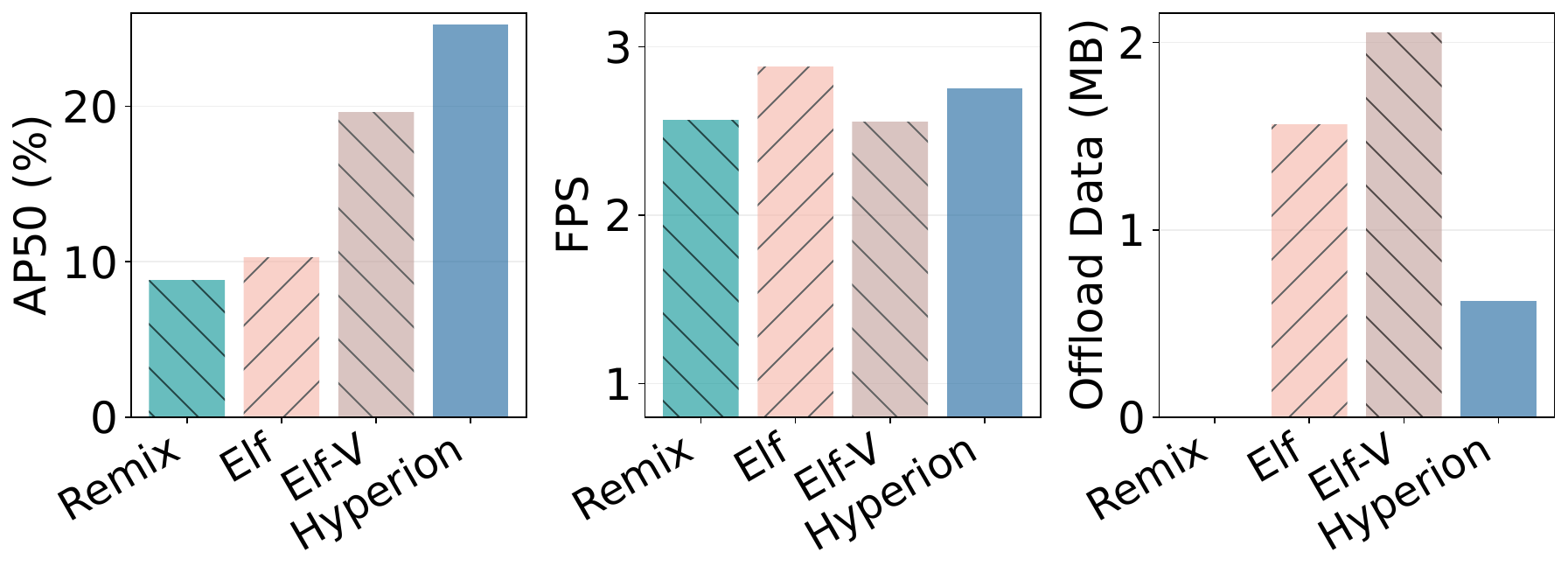}
    }
    \caption{Comparison of Hyperion and baselines in terms of accuracy, frame processing rate, and offload data across different network conditions.}
    \label{fig:gen-res}
\end{figure}

\subsection{Hyperion Performance} \label{sec:Overall Performance}

\textbf{Q1: Overall Performance.} We evaluate the overall performance of Hyperion under different network conditions, as shown in Fig.~\ref{fig:gen-res}. Hyperion consistently achieves a better trade-off between accuracy and frame processing rate than all baselines across both 4G and 5G scenarios, where the average bandwidth is 47.9 Mbps and 93.2 Mbps, respectively.
Compared to Remix and both versions of Elf, Hyperion improves accuracy by 5.6\%–20.2\%, while maintaining a competitive or superior frame processing rate. In terms of communication efficiency, Hyperion reduces the offloaded data volume to 24\%–39.8\% of that required by Elf and Elf-V.
Under similar mean E2E latency, Hyperion consistently outperforms the baselines in accuracy. For example, under 4G conditions, Hyperion reaches an AP50 of 22.8\% at 385.4 ms, significantly higher than Remix (8.8\% at 389.8 ms) and Elf (6.3\% at 387.9 ms). 
Meanwhile, Hyperion surpasses the best-performing baseline in accuracy while maintaining a lower latency. Under 5G, Elf-V achieves an AP50 of 19.6\% at 391.8 ms, while Hyperion achieves a higher AP50 of 25.3\% with a reduced latency of 362.9 ms.
Note that the relatively low accuracy scores are due to the inherent difficulty of the dataset \cite{wang2020panda}.
These results show Hyperion effectively balances accuracy, frame processing rate, and communication cost, outperforming existing SOTA baselines.

\begin{figure}[t]
\centerline{\includegraphics[width=0.50\textwidth]{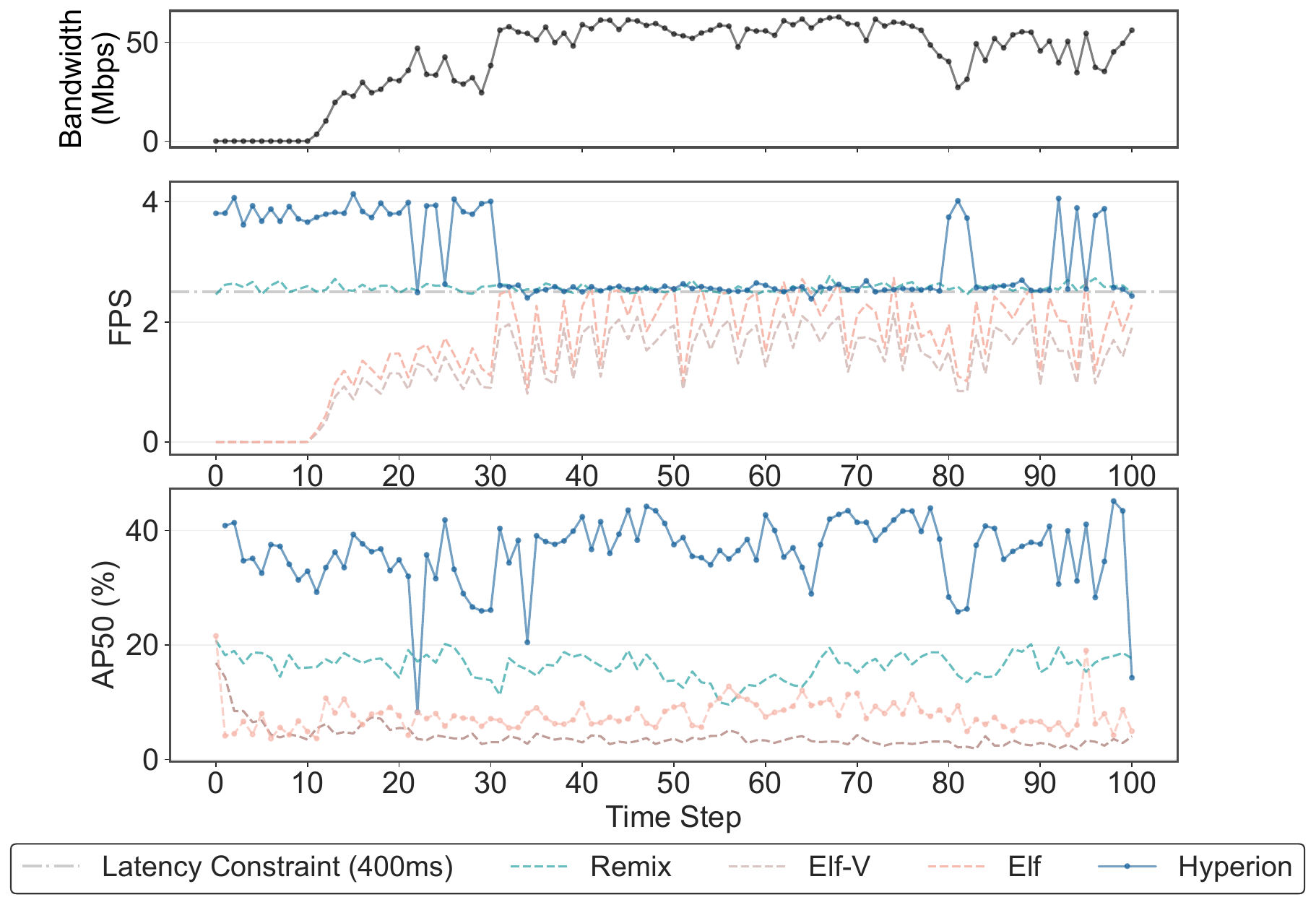} }
\caption{Runtime Analysis in a dynamic 4G LTE network trace.}
\label{fig:reason}
\end{figure}

{\bf Q2: Impact of Network Condition.} In Fig. \ref{fig:reason}, the top figure presents bandwidth traces of 4G LTE (LTE Driving, Run 8 trace) networks over a specific duration. The middle and bottom panels illustrate the corresponding frame processing rate and inference accuracy for each method over time.
In scenarios with low bandwidth(e.g., time step $\leq$ 11), Elfs experiences significant performance degradation due to high transmission latency, while Remix maintains low accuracy by relying on lightweight traditional models (i.e., EfficientDet-4\cite{tan2020efficientdet}) to meet latency constraints. In contrast, Hyperion achieves both higher throughput and accuracy. 
This is because Hyperion adaptively reduces its reliance on cloud inference by executing a lightweight vision transformer locally. 
In scenarios with a relatively high bandwidth(e.g., time step $>$ 11), the communication latency decreases. Hyperion uses the saved latency to maintain accuracy by transmitting frames at a reasonable quality level for inference on a more complex cloud model.
Additionally, as network conditions improve more, communication latency is no longer the primary bottleneck. Hyperion chooses to send original frames to the cloud, achieving the highest possible accuracy.
Overall, Hyperion dynamically adjusts its inference strategy based on network conditions, effectively balancing latency and accuracy. This adaptability enables Hyperion to outperform baseline methods under varying network conditions consistently.

\begin{figure}[t]
\centerline{\includegraphics[width=0.5\textwidth]{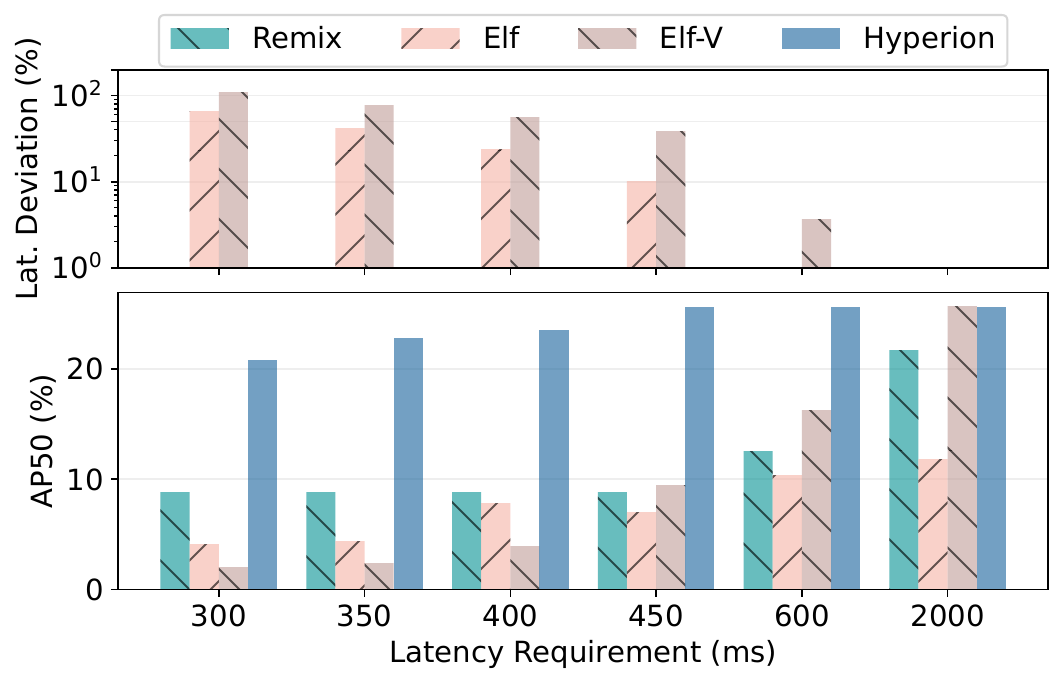} }
\caption{The inference accuracy and latency deviation ratio under varying latency requirements.}
\label{fig:var-latency}
\end{figure}

{\bf Q2: Impact of Latency Requirement.}
We next investigate the sensitivity to latency requirements. For brevity, we focus on the 4G LTE network conditions with a fixed uplink bandwidth of 47.9 Mbps.
Fig.  \ref{fig:var-latency} illustrates the accuracy and latency violation ratio for different methods under various latency constraints.
Hyperion consistently meets all latency constraints while achieving the highest accuracy across all latency requirements. Even under the strictest latency constraint (i.e., 300 ms), Hyperion achieves an accuracy of 20.79\%, which is 12.0\% higher than the second-best method.
Among baselines, Remix also satisfies all latency constraints but suffers significant performance drops under strict constraints(e.g., $\leq$ 450 ms). 
As the latency budget is relaxed to 2000 ms, Remix reaches an accuracy of 21.6\%, which is comparable with Hyperion. However, it incurs significant offline latency from generating a large number of partition plans. 
Elf and Elf-V exhibit high latency violation rates under tight constraints due to their large transmission volumes, which result in degraded accuracy. As the latency budget increases (e.g., to 650ms), both begin to satisfy the constraints consistently. Elf-V approaches Hyperion in accuracy, while Elf continues to underperform due to limitations of its traditional vision model. 
These results show that Hyperion effectively adapts to varying latency constraints while maintaining high inference accuracy.\\

\textbf{Q3: Ablation Study.}
We evaluate three variants of Hyperion. The first variant, V1, disables the importance scorer and directly uses Equation (2) as the importance score. The second variant, V2, replaces the adaptive transmission scheduler with a fixed policy, where patches from different classes are transmitted using fixed quality levels (i.e., $q_0 = 15, q_1 = 45, q_3 = 75$). 
The third variant, V3, replaces the cloud-device model ensembler by directly using the cloud results when they are available.
Compared to Hyperion, V1 increases the mean transmission latency by 9.7\% while maintaining similar accuracy. V2 results in a 4.9\% increase in mean E2E latency and an accuracy drop from 25.3\% to 10.2\%. V3 results in an accuracy drop of 0.4\% with similar mean E2E latency. This confirms the effectiveness of the key components of Hyperion.

\textbf{Q4: Overhead Analysis.}
We conducted an overhead analysis based on our prototype. The E2E latency of Hyperion mainly comes from four parts, including system overhead, data transmission, device computing, and cloud computing. The breakdown of their time consumption is illustrated in TABLE \ref{tab:overhead}. 
In real-world deployments, we tested Hyperion across WiFi 5, 5G, and 4G networks, with average bandwidths of 29.0 Mbps, 22.1 Mbps, and 6.3 Mbps, respectively.
To accommodate the network conditions, we set latency constraints of 500ms.
As can be seen, 
the overall system overhead is less than 3.5\%, indicating Hyperion is lightweight.

\section{Related Work}
\begin{table}[t]
    \centering
    \caption{The overhead analysis of Hyperion under different network conditions.}
            \begin{tabular}{c|c|c|c|c}
                \hline
                \multirow{2}{*}{\textbf{Network}} & \multicolumn{4}{c}{\textbf{Normalized Overhead (\%)}} \\ \cline{2-5} 
& \textbf{System} & \textbf{Device} & \textbf{Transmission} & \textbf{Cloud} \\
                \hline
                WiFi 5 & 1.8\% & 51.8\% & 37.0\% & 9.3\% \\
                \hline
                5G & 2.0\% & 56.6\% & 31.2\% & 10.2\% \\
                \hline
                4G & 3.5\% & 96.5\% & 0.00\% & 0.00\% \\
                \hline
            \end{tabular}
            \label{tab:overhead}
\end{table}
\textbf{Model-level High-resolution Inference Acceleration.}
Handling high-resolution images and videos as inputs in deep learning models poses significant challenges, particularly for vision transformers. The key difficulties arise from the self-attention mechanism in transformers, where the computational complexity grows quadratically with the input size, impacting inference speed and the number of parameters \cite{bakhtiarniaEfficientHighResolutionDeep2024a}.
A substantial body of research has focused on addressing these challenges by accelerating high-resolution vision transformers\cite{zhangMultiscaleVisionLongformer2021,yuanHrformerHighresolutionVision2021,guMultiscaleHighresolutionVision2022,yuGlancegazeVisionTransformer2021,chenSparsevitRevisitingActivation2023a,chenCrossvitCrossattentionMultiscale2021,yuMegabytePredictingMillionbyte2023}.
One line of work involves attention approximation \cite{zhangMultiscaleVisionLongformer2021,yuanHrformerHighresolutionVision2021,guMultiscaleHighresolutionVision2022,yuGlancegazeVisionTransformer2021,chenScalingVisionTransformers2022a,chenSparsevitRevisitingActivation2023a}.
Another line of work focuses on hierarchical processing \cite{chenScalingVisionTransformers2022a,chenCrossvitCrossattentionMultiscale2021,yuMegabytePredictingMillionbyte2023}. 
While much effort has been made to optimize vision transformers, existing solutions focus on designing more efficient models, often requiring architectural modifications that limit their compatibility with off-the-shelf models. We take a system approach to enable efficient Ultra-HD image processing on resource-limited devices.

\textbf{System-level High-resolution Inference Acceleration.}
Efficient high-resolution inference acceleration systems in traditional model architecture \cite{jiangFlexibleHighresolutionObject2021a,zhangElfAccelerateHighresolution2021a,yangFlexPatchFastAccurate2022b,shiAdaPyramidAdaptivePyramid2023, li2025riva}, such as CNNs and RNNs, have earned substantial research attention. 
These systems typically aim to extract and process specific ROIs from large input frames, as not all content in high-resolution videos is equally important. They manage ROIs either at the object-level\cite{zhangElfAccelerateHighresolution2021a,yangFlexPatchFastAccurate2022b} or at an even larger block-level\cite{jiangFlexibleHighresolutionObject2021a,shiAdaPyramidAdaptivePyramid2023}, resulting in severe system losses. 
Although recent works have focused on efficient vision transformer inference systems\cite{jiang2025janus, xu2023devit, ye2024galaxy}, these methods struggle to handle much higher communication and computation overheads in more challenging Ultra-HD scenarios.
In contrast to these studies, 
we design a novel inference system by exploiting the finer patch-level granularity inherent in vision transformers and realize superior performance. 
\section{Conclusion}\label{sec:Conclusion}
In this work, we present Hyperion, a cloud-device collaborative Ultra-HD inference system that enables low-latency vision transformer inference over dynamic networks. Hyperion builds on the inherent model characteristics identified in the vision transformer.
It integrates a collaboration-aware patch-level importance scorer to reduce transmission, a weighted small-large model ensembler to enhance accuracy, and a dynamic DP-based scheduler to balance latency and accuracy under dynamic network conditions. 
Extensive evaluations on real-world devices and network scenarios demonstrate that Hyperion significantly outperforms SOTA baseline methods in latency and accuracy. Our study presents a holistic system approach toward low-latency inference of vision transformer models on Ultra-HD vision input.

\bibliographystyle{IEEEtran}
\bibliography{IEEEabrv,references}

\end{document}